\documentclass{aa}  

\usepackage{graphicx}
\usepackage{color}
\usepackage{txfonts}

\usepackage{amstext}

\begin{document}

   \title{Stellar contributions to the line profiles of high-resolution transmission spectra of exoplanets}
   \titlerunning{Stellar contributions to the transmission spectra of exoplanets}
   \authorrunning{F. Borsa \& A. Zannoni}

   \author{
          F.~Borsa\inst{1}
          \and
          A.~Zannoni\inst{1,2}
          }

   \institute{INAF -- Osservatorio Astronomico di Brera, Via E. Bianchi 46, 23807 Merate (LC), Italy
   \and
Dipartimento di Fisica, Universit{\`a} degli Studi di Milano, via Celoria 16, 20133 Milano, Italy
             }
             \offprints{F.~Borsa\\ \email{francesco.borsa@inaf.it}}

   \date{Received ; accepted }

 
  \abstract
   {In-depth studies of exoplanetary atmospheres are starting to become reality. In order to unveil their properties in detail, we need spectra with a higher signal-to-noise ratio (S/N) and also more sophisticated analysis methods.}
   {With high-resolution spectrographs, we can not only detect the sodium feature in the atmosphere of exoplanets, but also characterize it by studying its line profile.
   After finding a clearly w-shaped sodium line-profile in the transmission spectrum of \object{HD~189733b}, we investigated the possible sources of contamination given by the star and tried to correct for these spurious deformations. 
   }
   {By analyzing the single transmission spectra of HD~189733b 
    in the wavelength space, we show that the main sodium signal that causes the absorption in the transmission spectrum is centered on the stellar rest frame.
   We concentrate on two main stellar effects that contaminate the exoplanetary transmission spectrum: center-to-limb variations (CLVs), and stellar rotation. 
   We show the effects on the line profile: while we correct for the CLV using simulated theoretical stellar spectra, we provide a new method, based directly on observational data, to correct for the Rossiter-McLaughlin contribution to the line profile of the retrieved transmission spectrum. }
   {We apply the corrections to the spectra of HD~189733b. Our analysis shows line profiles of the Na D lines in the transmission spectrum that are narrower than reported previously. The correction of the sodium D2 line, which is deeper than the D1 line, is probably still incomplete since the planetary radius is larger at this wavelength. A careful detrending from spurious stellar effects followed by an inspection in the velocity space is mandatory when studying the line profile of atmospheric features in the high-resolution transmission spectrum of exoplanets. Since the line profile is used to retrieve atmospheric properties, the resulting atmospheric parameters could be incorrectly estimated when the stellar contamination is not corrected for.
   Data with higher S/N coupled with improved atmospheric models will allow us to adapt the magnitude of the corrections of stellar effects in an iterative way.}
   {}


   \maketitle
%

\section{Introduction\label{sec:intro}}

Our knowledge of exoplanetary atmospheres has been growing exponentially in the past years as we pass from an era of exoplanet 
detection to an era of exoplanet characterization.
We are able to study the atmospheres of transiting and non-transiting exoplanets, provided that they orbit stars that are bright enough 
 \citep[see, e.g.,][]{2015PASP..127..941C,2016SSRv..205..285M,2017JGRE..122...53D}, but transiting planets provide more investigation possibilities
by far. 
When a planet transits, however, one effect has to be considered: the symmetry of the stellar disk emission is broken. If the star rotates, the symmetry is broken 
at the radial velocity (RV) of the area that is obscured by the orbital trajectory of the planet along the stellar disk.
As a consequence, the profile of the stellar lines is modified, as is the measured RV of the star; 
this effect (the Rossiter-McLaughlin effect, RM) 
can give insights into the inclination angle between the planetary orbital plane and the stellar rotational axis.

If a transiting exoplanet atmosphere hosts atomic or molecular species, the apparent radius of the planet will be larger at the 
specific wavelength of the spectral signatures of these species \citep[][]{2000ApJ...537..916S,2001ApJ...553.1006B}. The resulting wavelength-dependent depth of the transit
is the transmission spectrum of the exoplanet.
With high-resolution spectrographs, we can directly detect atomic features in the atmosphere of exoplanets. 
Sodium and potassium are among the most interesting elements because their cross sections are large. Because it is in the wavelength range of most optical spectrographs,
sodium is most frequently studied.
The first ground-based detection of sodium in a transmission spectrum was reported by \citet{2008ApJ...673L..87R}.
Recently, \citet[][hereafter W15]{wyttenbach} showed that sodium can not only be detected, but also characterized with the help of the high resolution of HARPS.
At a resolving power of R$\sim$100,000, it is possible to investigate high-altitude zones of exoplanetary atmospheres with lower pressures.
Studying the line profile of the planetary sodium absorption, we can constrain many aspects of its atmosphere \citep[e.g.,][]{2015ApJ...803L...9H,2017arXiv170909678P}.

When transmission spectra are reconstructed in the rest frame of the planet, the assumption of considering the 
RM to be symmetric and to cancel
out is no longer valid. It adds a spurious signal of stellar origin to the transmission spectrum, even 
with a perfect phase coverage \citep{louden,2016ApJ...817..106B}.
Center-to-limb variations (CLVs) of strong stellar lines 
can also introduce spurious signals of stellar origin that are due to planetary
occultation of different parts of the stellar disk  \citep{2015A&A...582A..51C,khala,2017arXiv170307585Y}. 

Using the same dataset and the same analysis method as W15, \citet{2016MNRAS.462.1012B} showed that the absorption is not recovered
for the sodium line alone, but also for H-$\alpha$ and CaII. These elements are not expected to be present in the atmosphere of the exoplanet. 
The authors also found evidence that the absorption is best recovered in the stellar rest frame, which challenges the detection of the planetary atmosphere and suggests that stellar activity might be a possible cause of the detection.

\citet[][hereafter W17]{2017arXiv170200448W} detected sodium absorption in WASP-49b using the same technique as presented in W15: this time, 
they took into account possible sources of contamination. They
studied a non-active star and analyzed the possible influence of the 
RM (which was not detected in the data).
\citet{2017A&A...608A.135C} also identified sodium absorption in WASP-69b with the same method, and also introduced the correction for telluric sodium using the 
second fiber of the HARPS-N spectrograph.

\

In this paper we analyze the effect that CLVs and RM have on the retrieved line profile of
the Na D lines in the transmission spectrum of exoplanets, focusing on HD~189733b. 
We show that a detrending of these effects is fundamental
in order to retrieve correct exoplanetary atmospheric parameters, in particular in the context of new-generation spectrographs, such as ESPRESSO.


\section{Data sample\label{sec:data_sample}}

We analyzed the HARPS datasets of three transits of HD~189733b as presented in W15, as this planet allows us to
use spectra with high signal-to-noise ratio (S/N).
We refer to that paper for observations logs, number of exposures, and exposure times for each transit.
Table~\ref{tabParameters} lists the stellar and orbital parameters we used in this work for the analysis of HD~189733b.

\begin{table}
\begin{center}
\caption{Stellar and transit parameters adopted in this work.}
\label{tabParameters}
\footnotesize
\begin{tabular}{ccc}
 \hline\hline
 \noalign{\smallskip}
 Parameter & Value & Reference\\
 \noalign{\smallskip}
 \hline
\noalign{\smallskip}
\multicolumn{3}{c}{HD~189733}\\
\noalign{\smallskip}
\hline
\noalign{\smallskip}
\multicolumn{3}{c}{\it Stellar parameters}\\
\noalign{\smallskip}
T [K]&   $4875\pm43$ & \citet{boyajian}\\
log g &   $4.56\pm0.03$ & \citet{boyajian}\\
 Fe/H &   $-0.03\pm0.08$ & \citet{torres}\\
\noalign{\smallskip}
\multicolumn{3}{c}{\it Transit parameters}\\
\noalign{\smallskip}
Period [days]&   $2.21857312$ & \citet{triaud}\\
$T_0$ [BJD] &   $2453988.80339$ & \citet{triaud}\\  
$R_{\rm p}$/$R_{\rm s}$ &   $0.1581$ &  \citet{triaud}\\
$R_{\rm s}/a$ &   $0.1142$ & \citet{triaud}\\
$e$ &   $0.0$ & assumed\\
$i$ [degrees]&   $85.508$ & \citet{triaud}\\
\noalign{\smallskip}
 \hline
\end{tabular}
\end{center}
\end{table}


\section{Data analysis\label{sec:data}}

We describe in this section the main procedures we used to extract the planetary transmission spectrum and to correct it for the stellar contamination.
We focused on the sodium signal, and in particular, on the Na D1-D2 lines (5895.92 and 5889.95 \AA, respectively). 

\begin{figure}[!ht]
\centering
\includegraphics[width=\linewidth]{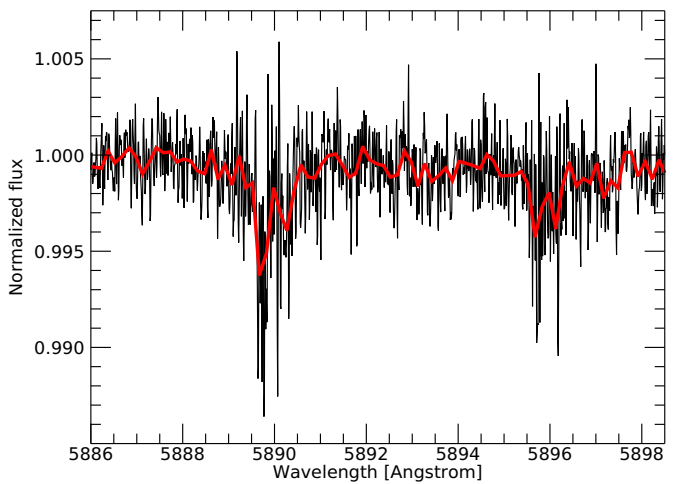}
\caption{
Sodium lines in the transmission spectrum of HD~189733b without correction for stellar effects. 
In red we overplot the binned spectrum, which shows the w-shaped profile of the sodium lines for illustration purposes.
}
\label{fig:W}
\end{figure}

\subsection{Transmission spectroscopy\label{sec:transmission}}
We performed transmission spectroscopy following the method of W15, which we briefly recall here.
Since HARPS is a highly stabilized spectrograph, we can consider any instrumental wavelength drift negligible.
The following steps were performed independently for each transit.
First we shifted the spectra to the null stellar radial velocity rest frame (i.e., at the velocity of the transit center) using a Keplerian model
of the system.
Then we normalized each spectrum by dividing by the average flux computed in the intervals 5868.5-5871.0 and 5910.8-5911.3 \AA\ close to the sodium D lines.


The problem of correcting telluric lines, especially in the core of strong stellar lines, has been a debated issue \citep[e.g.,][]{2005A&A...437..765H,2006A&A...448.1149H}. 
Using the out-of-transit spectra alone (except for night 2, as in W15), we built a telluric reference spectrum $T(\lambda)$ by means of  
a linear correlation between the logarithm of the normalized flux and the airmass \citep[][]{2008A&A...487..357S,2010A&A...523A..57V,2013A&A...557A..56A}. 
Using the scaling relation between airmass and telluric line strength, we rescaled all the spectra as if they had been observed at the airmass 
corresponding to that at the transit center. 
Since both the telluric and stellar rest frames during each analyzed transit observation move in the same direction and their variations are by far lower than the instrumental resolution, we performed the telluric correction in the stellar rest frame. In this way, we limited the number of interpolations. As a check, we divided the spectra by the exponential of the constant term of the linear regression (i.e., the normalized star) and performed it again in the telluric reference frame obtaining the same $T(\lambda)$ and the same final results.
\begin{figure*}[!h]
\centering
\includegraphics[width=\linewidth]{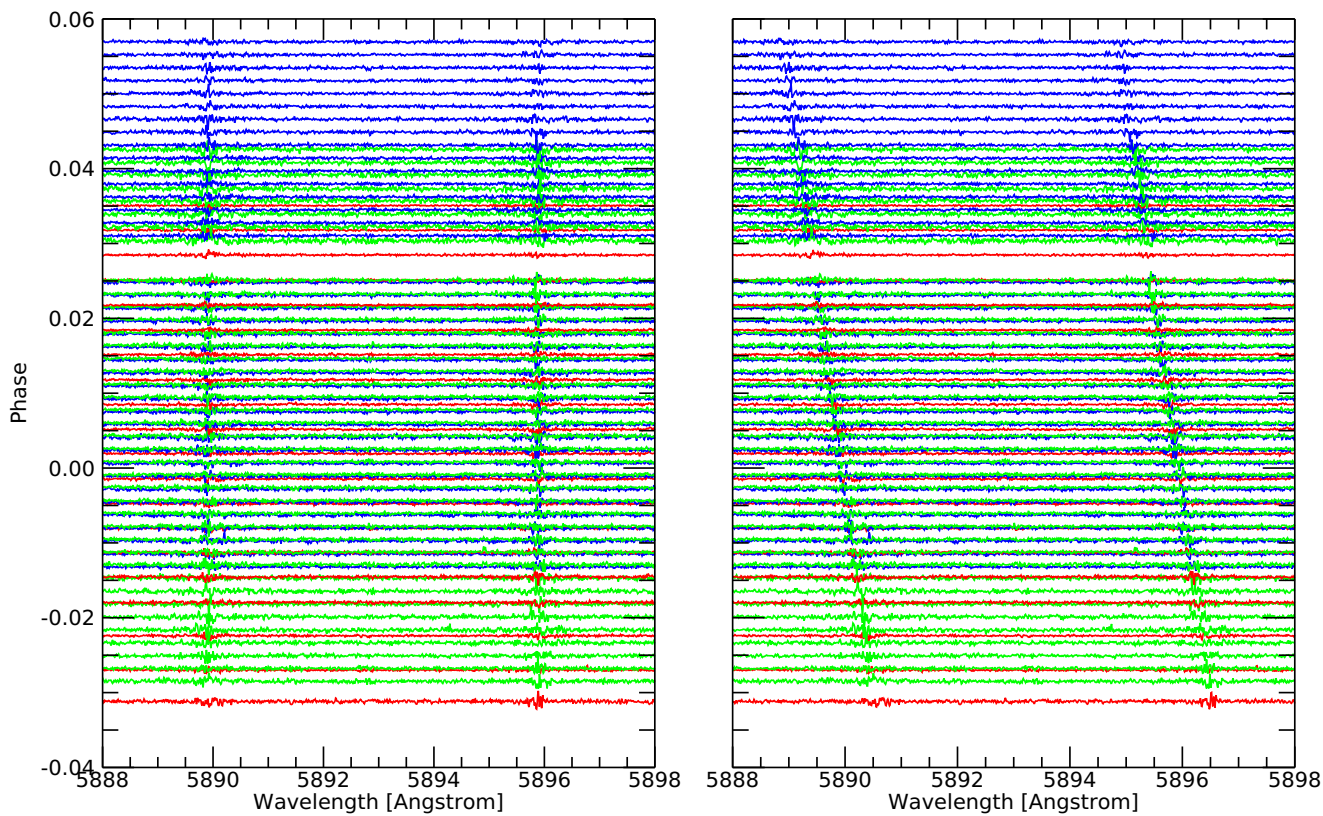}
\caption{
{\it (left)} Single residual spectra around the sodium D lines of HD~189733b aligned in the stellar reference frame. {\it (right)} Same as in the left panel, but aligned in the planetary rest frame. The main contribution, at least in terms of noise, is stationary in the stellar reference frame and moving in the planetary one. Different colors correspond to different transits. 
}
\label{fig:stellar_vs_planet}
\end{figure*}
Applying this method of telluric correction, we note that in correspondence of the sodium lines 
(and in general in correspondence of strong stellar lines because of the low S/N), the telluric reference spectrum $T(\lambda)$ presents some spurious features. 
Analyzing this effect in detail is beyond the scope of this work. We note that it might lead to possible shifts in wavelength of the signals when analyzing the final transmission spectrum line profiles as Gaussian fits, if the noise presents some random peaks that through the telluric correction might be propagated throughout the analysis.

 We then created a master stellar spectrum $M_{\rm star}$ by averaging all the out-of-transit spectra, and divided all the single spectra by this $M_{\rm star}$ to create the residual spectra $S_{\rm res}$.
To overcome intra-night atmospheric instabilities, we also performed the secondary telluric correction on the $S_{\rm res}$ spectra
to remove possible residuals, with a linear fit between $T(\lambda)$ and the spectra. 
After this, every $S_{\rm res}$ was shifted for the theoretical planetary RV and for the systemic velocity of the system, that is to say, we placed the spectra in the planetary reference frame. Here we expect to detect the exoplanetary atmospheric signal centered at the laboratory wavelengths of the sodium doublet. All the in-transit $S_{\rm res}$ can then be summed to create the average transmission spectrum for each single transit.

\begin{figure}[!ht]
\centering
\includegraphics[width=\linewidth]{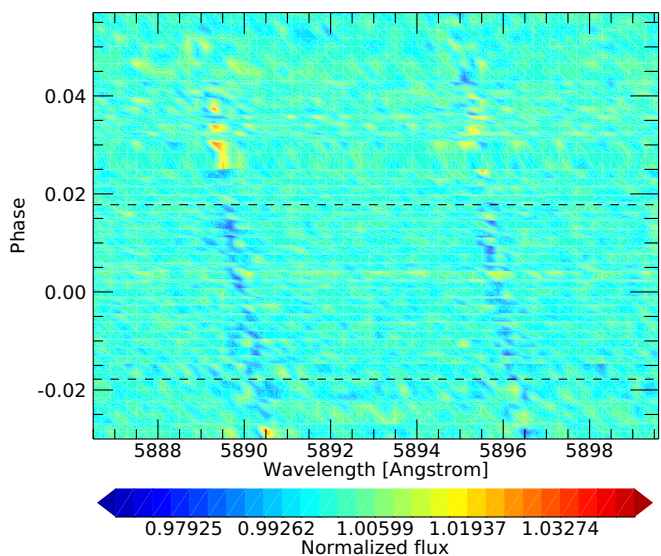}
\caption{Same as Fig.~\ref{fig:stellar_vs_planet}, binned in wavelength and shown in tomography in the planetary rest frame.
An absorption is present during the transit, whose limits are marked with dashed lines.
}
\label{fig:stellar_vs_planet_tomo}
\end{figure}

We noted a peculiar w-shaped line-profile of the Na D lines in the averaged transmission spectrum of the three transits (Fig.~\ref{fig:W}), therefore we searched for the possible causes of this distortion (see next sections). 
Instead of concentrating only on the sum of the in-transit spectra, we also analyzed the tomography of the residual sodium signal 
in a similar way as was done in \citet{2016MNRAS.462.1012B}. We find evidence that the main contribution to the sodium feature in the 
transmission spectrum, at least in terms of noise, is centered on the stellar reference frame (Fig.~\ref{fig:stellar_vs_planet}).
This might be obvious, since the noise is higher at the center of strong stellar lines, but looking more deeply,  an absorption signal behind the noise that is still centered on the stellar reference frame is also evident \citep[as was claimed by][]{2016MNRAS.462.1012B}, and it appears during the transit (Fig.~\ref{fig:stellar_vs_planet_tomo}).

  \subsection{Center-to-limb variations\label{sec:clv}}

The stellar flux is not uniform across the stellar disk. This non-uniformity, dependent on the wavelength, is
usually described by the limb-darkening parameters when analyzing planetary transits.
Each stellar line shows a variation in its profile from the center to the limb of the stellar disk, and strong lines 
can show quite significant variations \citep[e.g.,][]{2015A&A...582A..51C}.

\citet{khala} analyzed the CLV effect on the sodium of HD~189733b in terms of the different 
limb-darkening coefficients between the continuum and the line cores, using the formalism of \citet{mandelagol} with different coefficients to model the transit light curve.
We modeled the CLV effect in the same way as presented in \citet[][hereafter Y17]{2017arXiv170307585Y}.

The stellar spectra are created using the tool Spectroscopy Made Easy \citep[SME,][]{2017A&A...597A..16P}, using the MARCS \citep{2008A&A...486..951G} 
stellar atmosphere model and the line list from the VALD database \citep{2015PhyS...90e4005R}. We simulated one spectrum for 
each of the 21 different $\mu$ values considered ($\mu=\cos{\theta}$, with $\theta$ the angle between the normal to the stellar surface and the considered line of sight),
using the values of Table~\ref{tabParameters} for the stellar parameters.
For the sake of uniformity with Y17, we modeled the star as a grid with a radius of 101 points, corresponding to bins of 0.01 R$_{\rm star}$. 
We assigned a spectrum to each point by interpolating between the $\mu$ values for which they were simulated using quadratic interpolation.
Then we modeled the transit of the planet, calculating a stellar spectrum for each orbital phase as the average of the non-obscured simulated spectra.
 
\begin{figure}[ht]
\centering
\includegraphics[width=\linewidth]{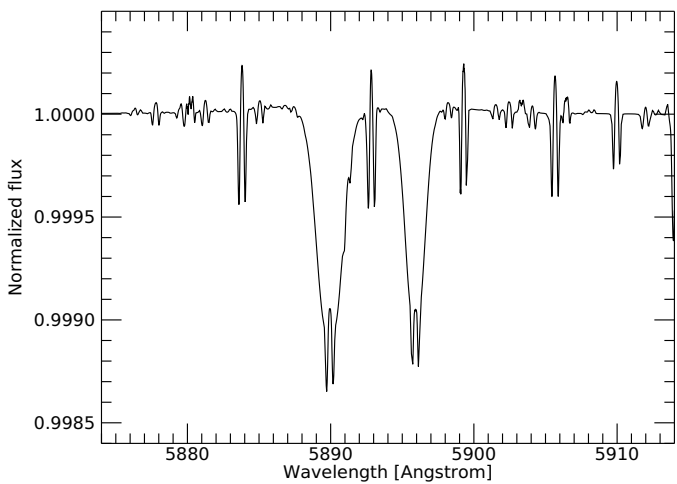}
\caption{
CLV effect for HD 189733 over an entire transit when shifted in the planetary rest frame. A w-shaped trace is given for each stellar line. 
For strong stellar lines, such as the sodium D-lines, a spurious absorption is also created, acting in the same way as a planetary atmospheric signal. 
}
\label{fig:clv_lineprof_189}
\end{figure}

Y17 used this method too to correct the sodium transit light curve of HD~189733b. 
We first verified that using their stellar and orbital parameters, we could obtain their same results after the CLV correction (planetary absorption depths within $1\sigma$ error bars). Then we studied the CLV modification of the line profiles in the transmission spectrum.
The CLV produces a stellar contribution that contaminates the signal of the exoplanetary atmosphere. Furthermore, we report that in the planetary rest frame, we detect a w-shaped distortion of the line profile that is caused by the CLV (Fig.~\ref{fig:clv_lineprof_189}). 
The w-shape occurs because the CLV effect contributes more to the transit phases in which the signal is blueshifted and redshifted (ingress and egress), 
and the impact is lower close to mid-transit time. 
The CLV contribution was calculated using models of the stellar atmosphere. Consequently, approximations of the stellar atmospheric structure may imply a non-perfect modeling of the CLV effect. For more accurate results,  3D star models are required (also pointed out by Y17), in particular because this effect is more prominent at the edge of the star.

  \subsection{Transmission CCF\label{sec:trans_ccf}}

Stellar rotation indeed modifies the line profiles during the transit of a planet, and it has effects on the retrieval of the transmission spectrum. 
This has been noted by \citet{louden}. They simulated this effect for HD~189733b, showing that it causes a w-shaped distortion in the transmission spectrum. A similar simulation was also presented in \citet{2017A&A...608A.135C}. W17 presented an in-depth analysis of the non-influence of this effect on the transmission spectrum of WASP-49b because the VsinI of the star is not detected: to reach this conclusion, they studied the transmission cross-correlation function (CCF).
The CCF is provided by the HARPS DRS pipeline and is obtained by comparing the stellar spectrum with a weighted line mask model \citep{2002A&A...388..632P}.
Roughly speaking, it reproduces the behavior of the average profile of the stellar lines.

\begin{figure}[ht]
\centering
\includegraphics[width=\linewidth]{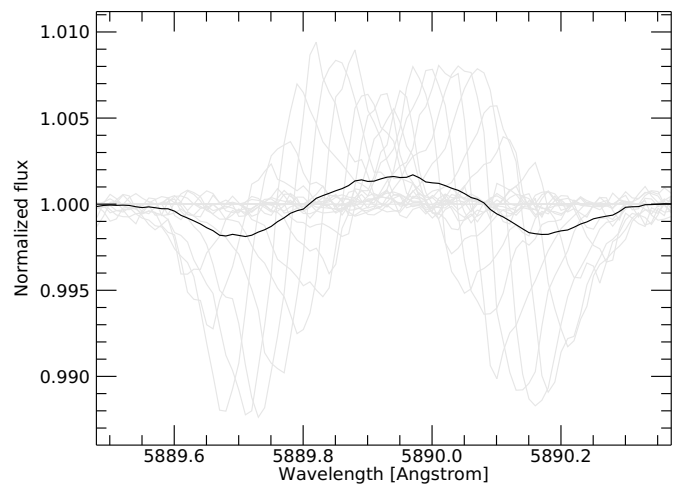}
\caption{
Consequence of the RM on the sodium D2 line when shifted in the planetary rest frame for the transit of night 3. 
Black line: global effect on the whole transit. Gray lines: effect on each single spectrum. 
}
\label{fig:ccf_trans}
\end{figure}

We introduce here a new method to correct for the rotational effect.
Unlike the correction for the CLV, which was derived from a model, this correction is obtained directly from the observational data.
Like in W17, we also took advantage of the CCF: studying it instead of the single line-profile allows us to 
provide a correction that is of higher quality.
We studied the behavior of the CCF profile during the transit by analyzing the residual transmission CCFs, which are the normalized CCFs shifted in the stellar rest frame and 
divided by the average normalized out-of-transit CCF.
We then rescaled these residual CCFs for the line-depth ratio between the average normalized out-of-transit CCF and the average out-of-transit line-profile of the line of interest (i.e., the Na D2-D1 lines in our case) in the stellar rest frame, switched from wavelength- to velocity-space.
When shifting all these rescaled transmission CCFs in the reference frame of the planet, we note that in their sum another w-shaped profile is evident (Fig.~\ref{fig:ccf_trans}), which arises because the planet obscures a rotating star (e.g., a consequence of the RM) with an angle of inclination close to $\lambda$=0.
We then divided the line of interest in the residual transmission spectra for the relative rescaled transmission CCF, converted in the wavelength-space. 
In this way, we corrected each single transmission spectrum for this effect for the sodium D1 and D2 lines.

The quality of the correction depends on the quality and on the temporal extension of the data: a good coverage of the out-of-transit phase ensures a higher S/N on the average out-of-transit CCF.
We also point out that this correction is an approximation: the CCF is a mean line-profile that covers all wavelength ranges of the spectrograph. The magnitude of the perfect correction will instead also depend on the wavelength if the apparent radius of the planet is dependent on it, as is the case here. 
This correction thus can account only for the white-light radius of the planet (see further discussion of this in Sect.~\ref{sec:discussion}).


\section{HD~189733b\label{sec:hd189733}}

\begin{figure}[!ht]
\centering
\includegraphics[width=\linewidth]{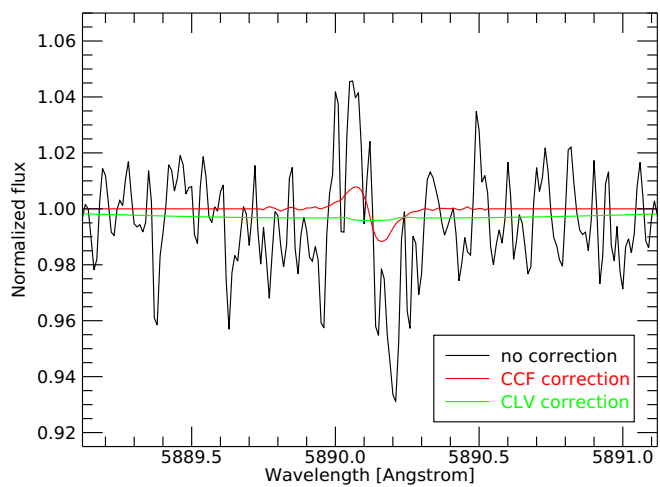}
\caption{ 
Example of a single transmission spectrum of HD~189733b, zoomed on the sodium D2 line, with the corrections of CLV and RM overimposed.
The RM correction in particular looks underestimated, as discussed in Sect.~\ref{sec:discussion}.
}
\label{fig:sodium_single_corrected_hd189}
\end{figure}

\begin{figure*}[!ht]
\centering
\includegraphics[width=\linewidth]{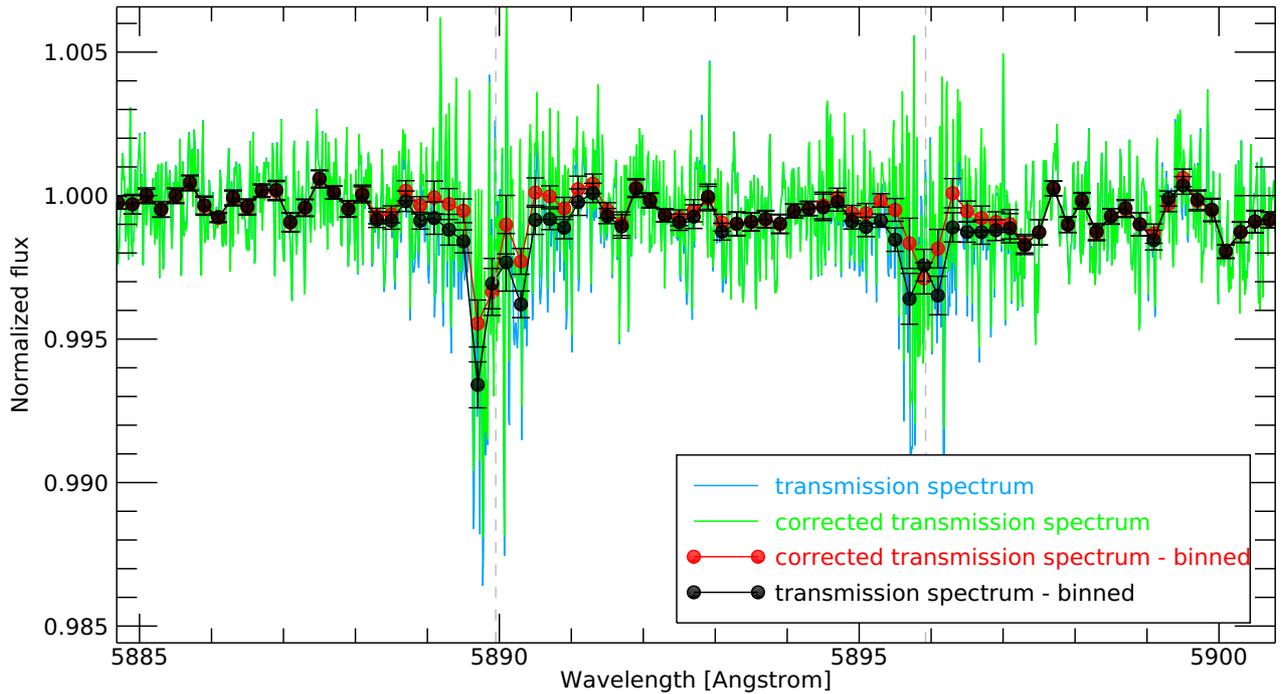}
\caption{ 
Atmospheric sodium line profiles of HD~189733b in the planetary rest frame before and after the correction for stellar effects. The corrected profiles are narrower. 
The w-shaped profile disappears after the corrections for the D1 line, while it remains for the deeper D2 line.
Vertical dashed lines represent the rest frame of the sodium D lines.
}
\label{fig:sodium_corrected_hd189}
\end{figure*}

We show here the detrending of the effects CLV and RM from the line profile of the sodium D1 and D2 lines of the transmission spectrum of HD~189733b.
Fig.~\ref{fig:sodium_single_corrected_hd189} shows an example of the corrections applied to a single transmission spectrum for the D2 line.
The line profile of the sodium D2 and D1 lines in the global average transmission spectrum of HD~189733b, before and after the corrections, is shown in Fig.~\ref{fig:sodium_corrected_hd189}.
While the D1 line after the correction loses the w-shaped profile, this happens only partially for the profile of the D2 line. We discuss possible causes of this in Sect~\ref{sec:discussion}.

\begin{figure*}[!ht]
\centering
\includegraphics[width=\linewidth]{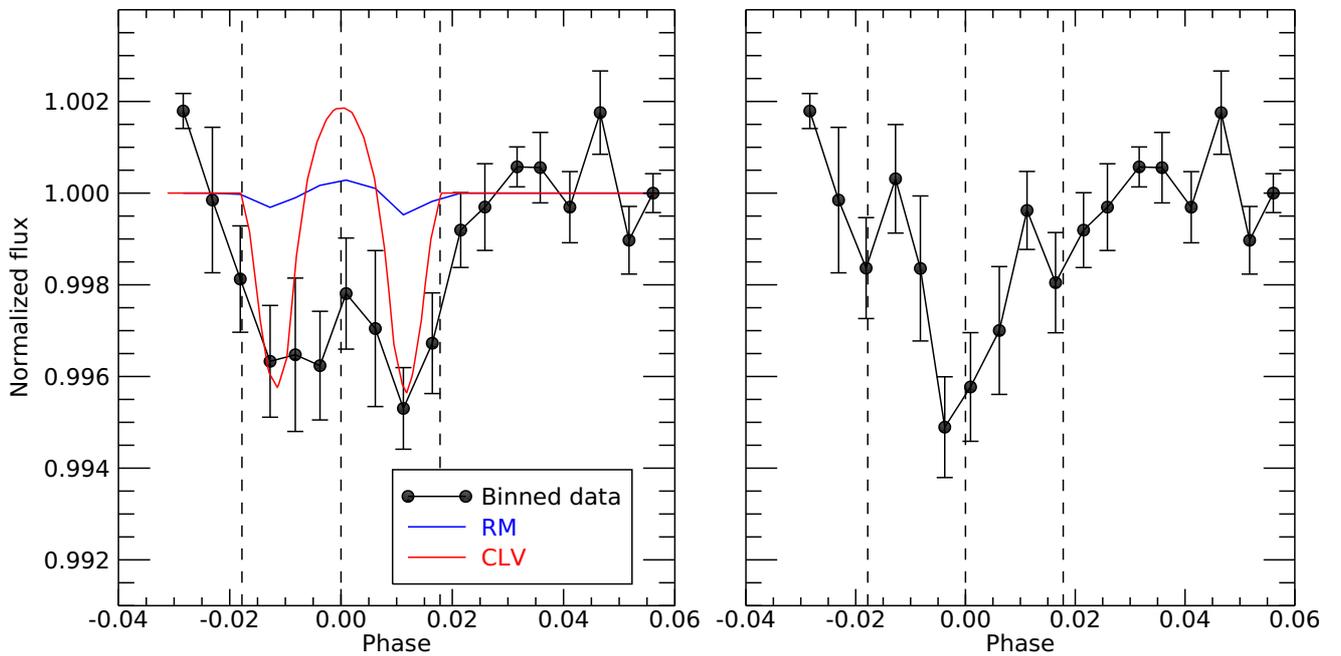}
\caption{
Transmission light curve of HD~189733b in the planetary rest frame from the three transits, centered on the sodium D2 and D1 lines (averaged), with bandpasses of 0.75 Angstrom. Vertical dashed lines represent the beginning, middle, and end of the transit based on ephemerides. {\it (left)} Before the corrections for stellar effects, which are overimposed. {\it (right)} After the correction for CLV and RM related effects.
}
\label{fig:tlc_189}
\end{figure*}

We also show in Fig.~\ref{fig:tlc_189} the transmission light curve calculated in bands of 0.75 Angstrom centered on the sodium doublet in the planetary rest frame. This has previously been corrected for the CLV effect in the literature (see Sect.~\ref{sec:intro}), but not for the RM. 


\section{Discussion\label{sec:discussion}}

The main contributions in the raw transmission spectrum of HD~189733b, before any correction for stellar effects, 
are centered on the stellar rest frame (Fig.~\ref{fig:stellar_vs_planet}). 
At first glance, this might be thought to be due to noise, because the core of deep lines such as the sodium Na D 
lines is by far more noisy than the continuum. There is also clear evidence of absorption during the transit in the stellar rest frame, however (Fig.~\ref{fig:stellar_vs_planet_tomo}).


\
 We removed CLV and RM effects in the exoplanetary transmission spectrum of HD~189733b, that is, in the resolved profile of the sodium D1 and D2 lines. 
 The correction for the CLV effect was made starting from stellar models, while to remove the rotational effect, we started from real data using the CCFs.
 In our method, we correct every single spectrum for the stellar effects, not only the summed transmission spectrum. 
 For the case of HD~189733b, the main spurious contribution given by the stellar effects in the retrieved line profiles
 is clearly the contribution by stellar rotation (see Fig.~\ref{fig:sodium_single_corrected_hd189} for an example). The correction for the CLV effect is by far less important in the line profile. However, we note that this is specific 
 for the star analyzed. Beyond the planet-to-star radius ratio, the magnitude of the CLV effect depends on the stellar temperature (Y17), and that 
 of the RM on the VsinI of the star.
 After removing these two stellar contributions, the line profile of the D1 line no longer shows a w-shaped distortion 
 (Fig.~\ref{fig:sodium_corrected_hd189}).
 For the D2 line, however, this distortion is still present.
 The absorption of the D2 line has often been found to be much larger than the one of the D1 \citep[e.g., ][]{2012MNRAS.422.2477H,khala,2017A&A...608A.135C}.
  We hypothesize that the non-cancellation of the w-shape is due to a non-perfect correction for the stellar effects. The CLV and RM corrections are both proportional to the planet-to-star radius ratio. If for a particular wavelength this is greater than the average one, as expected for the D2 line in the case we analyzed, the correction to be applied should be more important. 
 The magnitude of this overcorrection is not known a priori, however.
 We argue that for data with a higher S/N than currently available (S/N $\ge 3$ for each single residual spectrum), coupling the spectra
 with atmospheric models could allow us to adapt the magnitude of the corrections to the data in an iterative way.

In the transmission light curve centered on the sodium doublet, in contrast to what happens for the line profiles, the stellar contribution is higher for the CLV than for the rotation (see Fig.~\ref{fig:tlc_189} for a 0.75 Angstrom bandpass). The reason is that in the transmission light curve we integrate over a wide range of wavelengths. For the case of HD~189733b, the effects of RM are centered very closely to the center of the line, while for the CLV they are weaker in magnitude but persistent in a much greater wavelength range.

\

\begin{figure}[h]
\centering
\includegraphics[width=\linewidth]{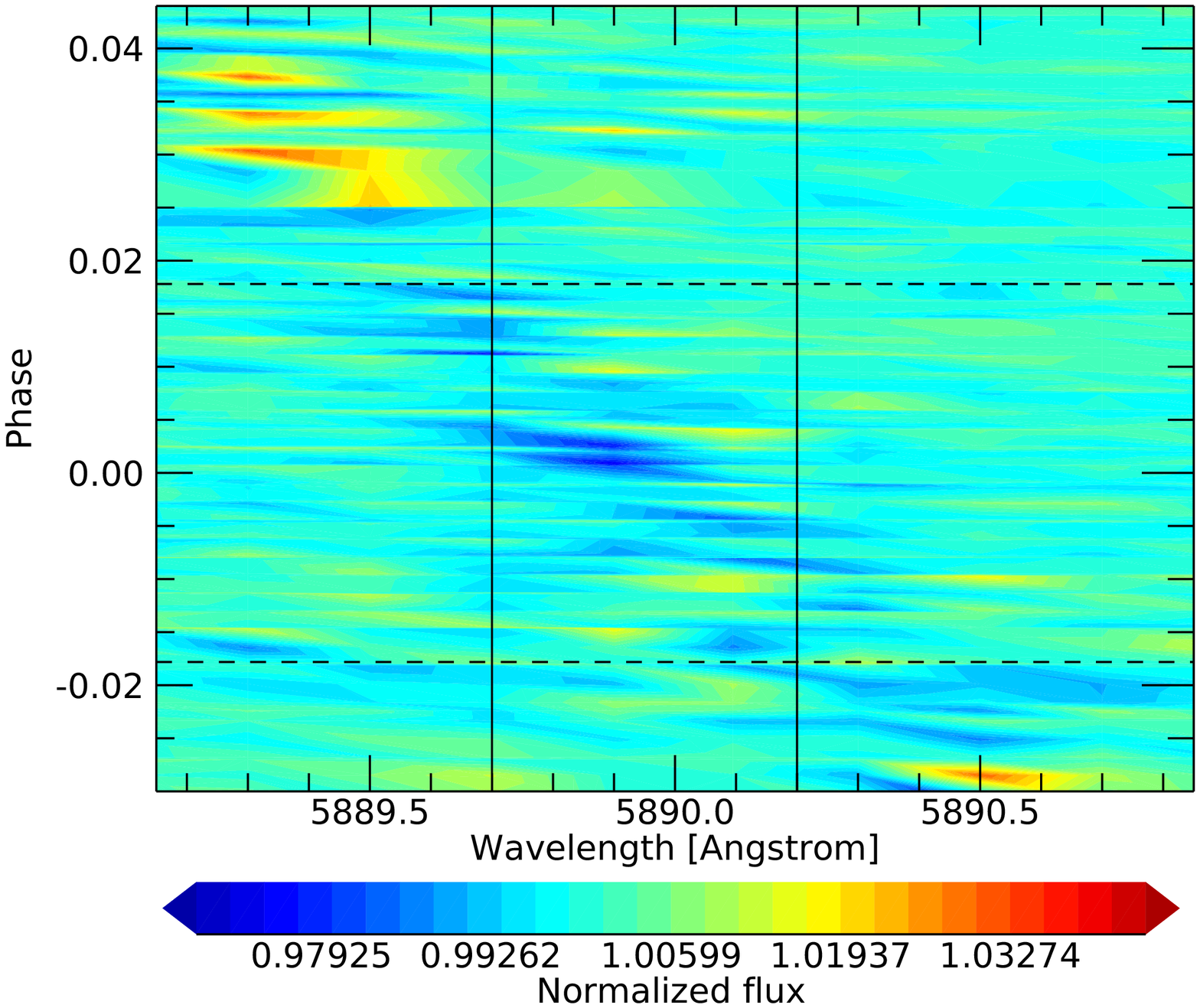}
\caption{ 
Tomography of the corrected sodium D2 line absorption. 
Horizontal dashed lines represent the phases of transit ingress and egress.
Vertical lines represent a 0.5 Angstrom passband centered on the D2 line: the transit inside this passband is shorter than the transit from ephemerides because the main signal is still in the stellar rest frame.
}
\label{fig:spiegazione}
\end{figure}

We note that after applying the corrections, the transit duration centered on the sodium D lines with 0.75 Angstrom bandpasses is shorter than the planetary transit in the transmission light curve (Fig.~\ref{fig:tlc_189}).
One possible interpretation of this effect is an imperfect correction for the stellar contribution.
Fig.~\ref{fig:spiegazione} shows that it is easy to understand that narrowing the bandwidth of the integration centered on the sodium line will cause a shorter transit duration if the main absorption is not centered on the planetary, but on the stellar rest frame.

\begin{figure}[h]
\centering
\includegraphics[width=\linewidth]{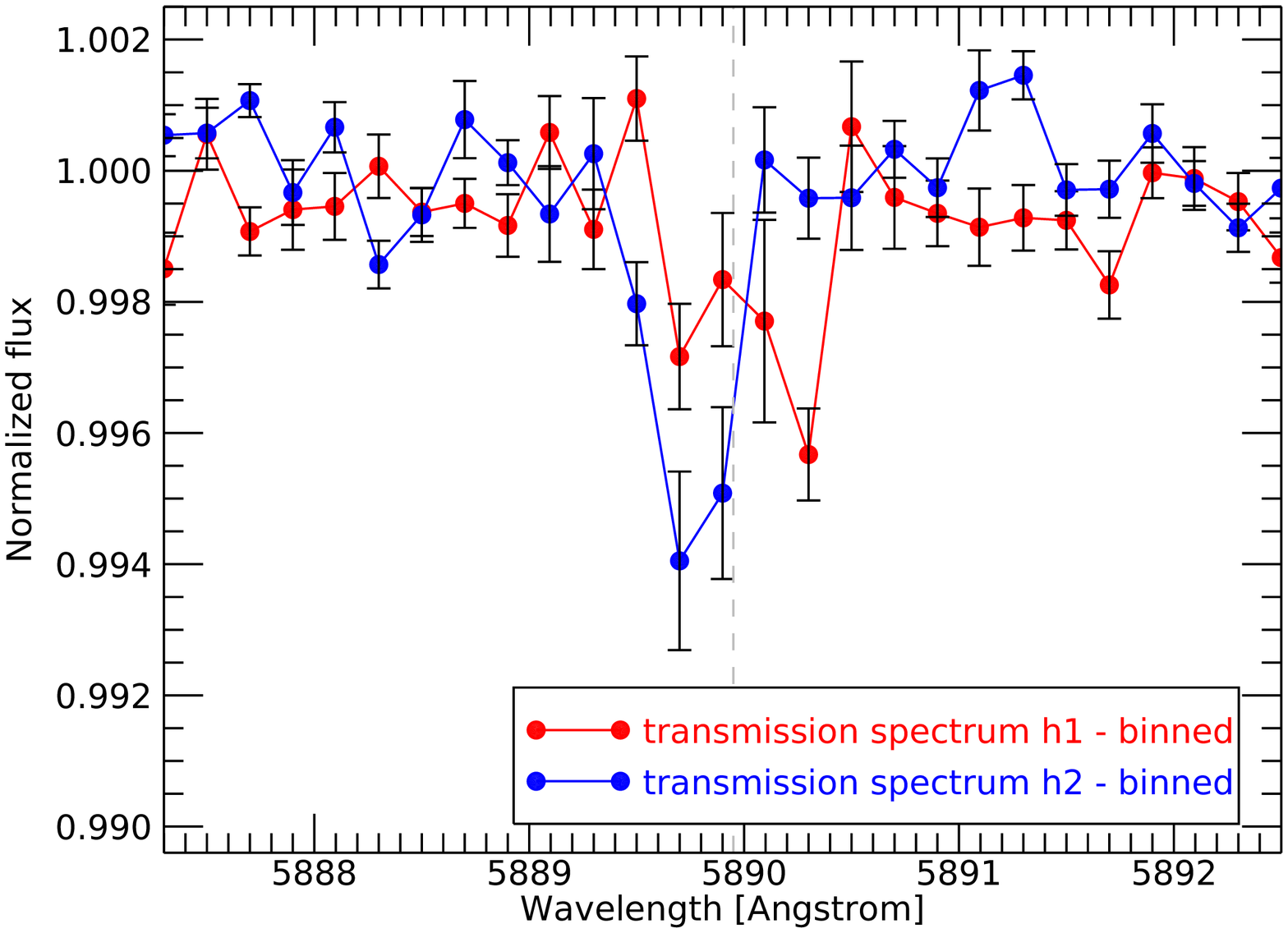}
\caption{ 
Transmission spectrum of the sodium D2 line, as calculated in the first (h1) and second half (h2) of the transit.
The two absorption excesses are redshifted and blueshifted with respect to the planetary rest frame, respectively.
The vertical dashed line shows the sodium D2 rest frame.
}
\label{fig:duemeta}
\end{figure}

\begin{figure}[h]
\centering
\includegraphics[width=\linewidth]{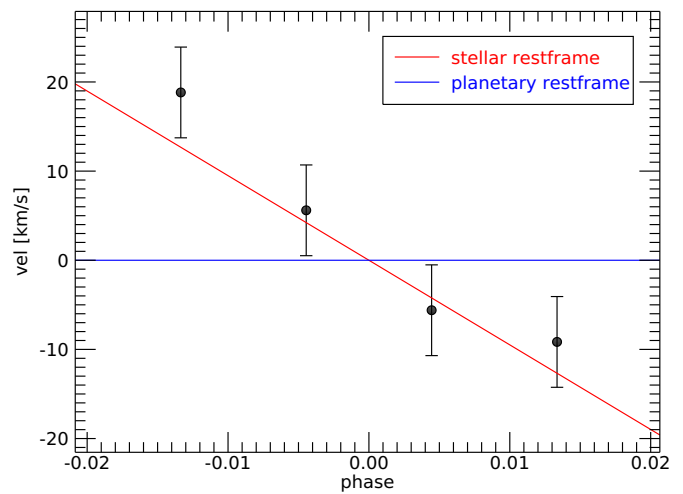}
\caption{ 
Velocity of the maximum absorption of the sodium D2 line in the planetary rest frame as it varies during the transit.
Error bars are taken as the width of the bin (0.2 Angstrom).
The rest frame of the planet (blue line) and the rest frame of the star (red line) are overplotted.
}
\label{fig:quattroparti}
\end{figure}

We further investigated the RV system on which the main absorption is centered by examining the transmission spectrum of each half of the transit (Fig.~\ref{fig:duemeta}). 
We analyzed the sodium D2 line because the S/N for the D1 line is not sufficient.
Its absorption is clearly redshifted or blueshifted, depending on which half of the transit is considered.
We also tried to verify the rest frame of the sodium D2 absorption by dividing the transit into four parts and binning the resulting transmission spectra by a factor 20$\text{}$. Because of the low S/N, and thus the 
impossibility to perform a Gaussian fit, we considered as reference the wavelength bin of maximum absorption for each of the four transmission spectra. 
Fig.~\ref{fig:quattroparti} clearly shows that the velocity of the main absorption signal matches the stellar rest frame.

\section{Conclusions\label{sec:conclusions}}

We presented a method to mitigate the stellar contribution in the high-resolution transmission spectrum of exoplanets.
The stellar effects taken into account are the stellar rotation (RM) and the CLV, which both contaminate the exoplanetary transmission spectrum. 
While for the CLV we used stellar models, we proposed a correction for the RM that is based directly on observational data.

We applied these corrections to the transmission spectrum of HD~189733b because the S/N of the available in-transit high-resolution spectroscopy is the highest of the known exoplanets.
The retrieved profiles of the sodium Na D lines become narrower: we note that a less broadened line-profile would also better agree with 
theoretical models \citep[e.g., ][]{2017arXiv170909678P}.
We stress that our correction works for an average planetary radius: for wavelengths where the radius is expected to be larger, this is currently not sufficient to retrieve the correct absorption line profiles.

We provided further evidence that without any correction for the stellar effects, the absorption signal in the planetary transmission spectrum is centered on the stellar rest frame.
We do note that there is still a residual contribution in the stellar reference frame even after the corrections (Fig.~\ref{fig:spiegazione}), at least for the sodium D2 line (the S/N is too low in the D1 line to perform the same tests). 
We stress the importance of verifying the corrections of RM and CLV in velocity space: any imperfect correction, which will be relevant in the final transmission spectrum, will then be evident.
This star is also very active \citep[e.g.,][]{boisse}: any out-of-transit average spectrum can therefore not be completely representative of the status of the star, and this can influence the retrieval of the transmission spectra and of the transmission CCF.
While the effect of CLV on the line profile is not so important for HD~189733b, it could be more relevant for other planets, in particular those orbiting
M-type stars because the magnitude of the CLV depends on the stellar temperature.

The line profiles retrieved in the transmission spectrum are used to derive the properties of the exoplanetary atmospheres: a clear separation of planetary and stellar contribution is indispensable to correctly characterize exoplanets.
To better understand the stellar contamination and to correct the retrieved line profiles in the transmission spectra of exoplanets, spectra with a high S/N are fundamental.
High-resolution stabilized spectrographs on larger telescopes such as ESPRESSO \citep{2014AN....335....8P} or the project HIRES \citep{2016SPIE.9908E..23M} will certainly help, but
this is another reason for continuing the hunt for new exoplanets that transit bright stars with dedicated ground- \citep[e.g., NGTS; ][]{2013EPJWC..4713002W} and space-based \citep[e.g., PLATO; ][]{2014ExA....38..249R} surveys.


\begin{acknowledgements}

We thank the referee for their interesting and useful comments that helped improve the clarity of the paper.
Based on observations made with ESO Telescopes at the La Silla Paranal Observatory.
FB acknowledges financial support from INAF through the ''Progetti Premiali'' funding scheme of
the Italian Ministry of Education, University, and Research and through the ASI-INAF contract 2015-019-R.O.
AZ acknowledges support from Prof. Giuseppe Lodato during his Master thesis work.

\end{acknowledgements}


\end{document}